\documentclass[a4paper,11pt]{article}
\pdfoutput=1 % if your are submitting a pdflatex (i.e. if you have
\usepackage{jinstpub} % for details on the use of the package, please
\usepackage{lineno}
\title{\boldmath Numerical study of RF power coupling in fusion-relevant single- and multi-driver H$^-$ ion sources}

\author[a,1]{D. Zielke,\note{Corresponding author.}}
\author[a]{S. Briefi,}
\author[a,b]{U. Fantz}%\note{Also at Some University.}}
\affiliation[a]{Max Planck Institute for Plasma Physics, Garching, DE}
\affiliation[b]{Augsburg University, Augsburg, DE}
\emailAdd{dominikus.zielke@ipp.mpg.de}

\abstract{ITER's large and powerful neutral beam injection system is based on an ion source utilizing a modular concept, where eight cylindrical drivers are attached to one common expansion and extraction region. In each driver, a plasma is sustained via inductive coupling with powers of up to 100\,kW at a driving radio frequency (RF) of 1\,MHz to produce fusion-relevant hydrogen beams. These high powers impose great stress on the electric system. Recent measurements at the single-driver test bed BATMAN Upgrade showed that the RF power transfer efficiency $\eta$, which measures the ratio of power absorbed by plasma to total RF power, is only around 0.5, leaving room for optimization. In multi-driver test beds such as ELISE with four drivers $\eta$ is found to be even further decreased to around 0.4. To explain this difference, a previously validated self-consistent 2D RF power coupling fluid model is applied. For the same absorbed power per driver, the model shows virtually the same spatial distributions of plasma parameters and power absorption in single- and multi-driver sources. However, the coil current is slightly increased in the multi-driver model due to a changed spatial distribution of the magnetic RF field in the region surrounding the drivers. Typically, in multi-driver sources conductive shields are applied to cancel the electromagnetic interference between individual drivers. These shields are found to affect the spatial distribution of the RF fields more severely, the effect being highly dependent on the distance between the RF coil and the shield. In the case of the ELISE ion source a further decrease of $\eta$ is calculated by the model being in good agreement with experimental measurements.}


\proceeding{8$^{\text{th}}$ international symposium on Negative Ions, Beams and Sources - NIBS$'22$\\
  Oct 2 - 7, 2022\\
  Padova, Italy}

\begin{document}
\maketitle
\flushbottom
\section{Introduction}

\begin{figure}[ht]
	\centering
	\includegraphics[width=0.9\linewidth]{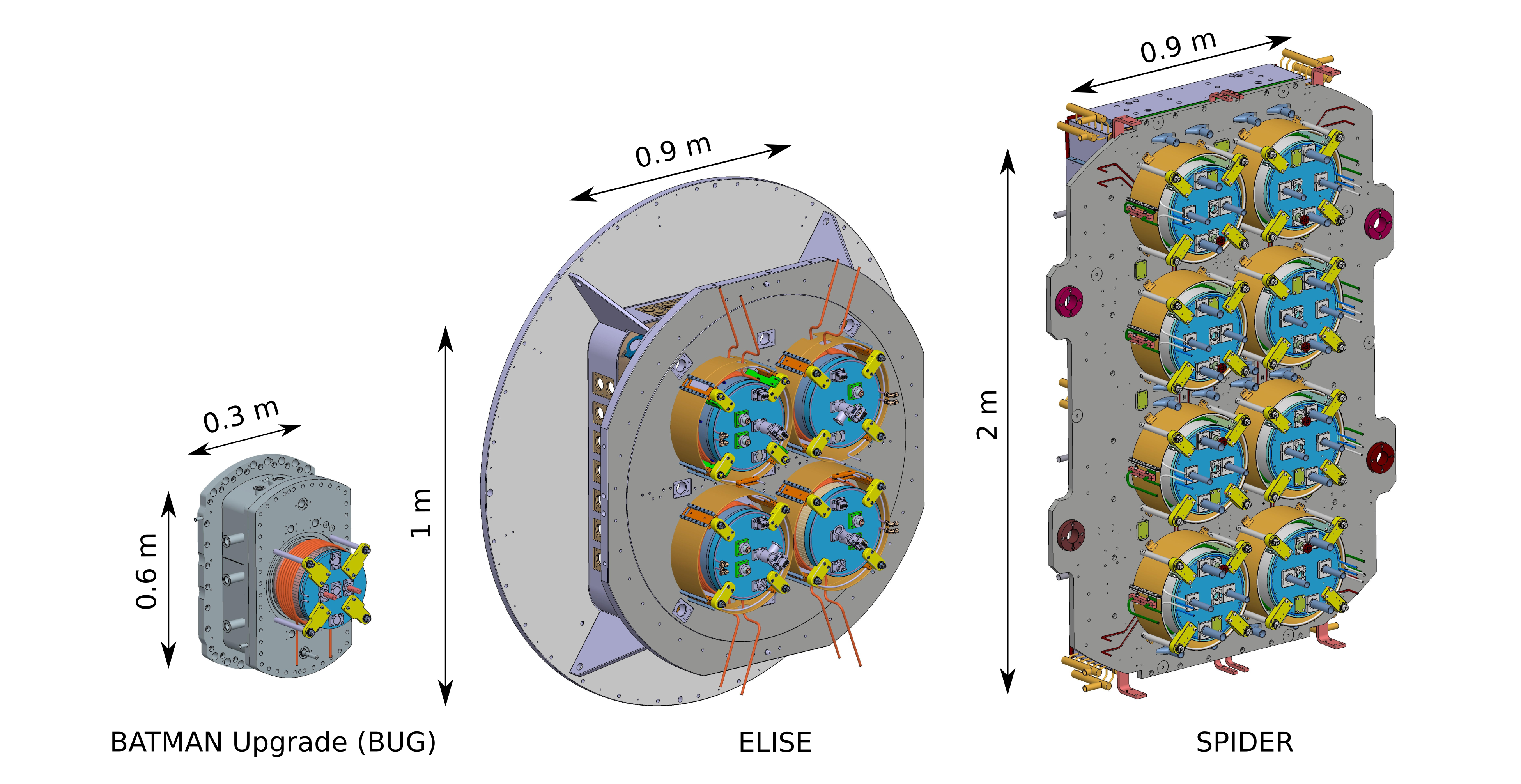}
	\caption{\label{fig:CAD_drivers}Size scaling of the modular radio frequency ion source concept: one driver at the BUG test bed (left), four drivers at the ELISE test bed (middle), and the full size ITER ion source at the SPIDER test bed with eight drivers (right). Figure based on Heinemann et al.~\cite{Heinemann_2017}.}%
\end{figure}

Neutral beam injection (NBI) systems as envisaged for the ITER experiment~\cite{ITER} produce large and powerful hydrogen or deuterium beams, which are injected into a magnetically confined high temperature plasma to heat it to fusion-relevant temperatures~\cite{Hemsworth_2017}. For this purpose, a modular ion source concept is foreseen, where multiple cylindrical drivers are attached to one rectangular expansion and extraction region~\cite{Heinemann_2017}. Figure~\ref{fig:CAD_drivers} shows the size scaling from the single-driver prototype ion source at the BATMAN Upgrade (BUG) test bed towards the multi-driver test beds ELISE with four drivers, and the full size ITER ion source at the SPIDER test bed with eight drivers~\cite{Heinemann_2017}.

RF coils with six to eight windings are wrapped around each driver and connected to RF generators via capacitive matching networks and transmission lines. At the BUG and ELISE test beds, matching transformers are used to separate the RF generators which are on ground potential from the ion sources, which are on a high negative potential of up to $-60\,\mathrm{kV}$. At the SPIDER and other ITER sources both the RF generators as well as the ion sources are on a high negative potential, wherefore no matching transformer is necessary. In the multi-driver test beds each two horizontal drivers are connected in series and supplied by one common RF power generator operating at a radio frequency (RF) of 1\,MHz. As shown in figure~\ref{fig:CAD_drivers}, one striking difference between the single- and multi-driver sources (besides that the BUG test bed is in air and all others are in vacuum vessels) is that in the latter case cylindrical conductive electromagnetic shields (EMS) are introduced around each RF coil. This was done to suppress the electrostatic and electromagnetic coupling between the drivers~\cite{Kraus_2012}.

In these ion sources negative hydrogen ions are produced mainly on the surface of a cesiated plasma facing grid in the expansion by conversion of impinging atoms and positive ions. Hence it is mandatory to sustain a hot and dense plasma in the drivers, where typical plasma densities and electron temperatures are around $10^{18}\,\mathrm{m}^{-3}$ and 10\,eV, respectively and the hydrogen atomic to molecular ratio is around 0.5. Up to 100\,kW of RF power per driver has to be delivered by the RF generators to achieve the ITER performance~\cite{Hemsworth_2017}. These high powers entail high currents and voltages in the RF coils and matching networks, rendering the system prone to RF breakdowns and failures. It has been determined experimentally at the prototype RF ion source of the BUG test bed that the fraction of generator power which is absorbed by the driver plasma does not exceed 0.6 in a single-driver ion source~\cite{Zielke_2021}. The considerable loss power was found to be distributed between Joule heating losses in the skin of the RF coil (25\%) and eddy current losses in the internal Faraday shield (75\%). The latter has to be present in RF ion sources to protect the dielectric cylindrical wall material from plasma sputtering and the plasma from impurities~\cite{Kraus_2001}.

To optimize the RF power coupling a numerical fluid-electromagnetic model was developed and benchmarked successfully~\cite{Zielke_2022}. The model calculates the RF fields and the macroscopic plasma quantities self-consistently in the operating regime of RF ion sources, i.e.\ at large power densities in the order of 10\,Wcm$^{-3}$, low RF of 1\,MHz and at the ITER required low gas pressure of 0.3\,Pa. Among the various investigated external parameters, increasing the axial driver length (from 17\,cm to 30\,cm) and driving frequency (from 1\,MHz to 2\,MHz) were identified to yield promising improvements for the RF power coupling, wherefore less losses in the Faraday shield and RF coil are to be expected~\cite{Zielke_2022_2}.

When compared to single-driver ion sources, the fraction of absorbed power in multi-driver ion sources is even further decreased. This has been experimentally observed at the ELISE test bed, where the fraction of absorbed power ranges around 0.4 ($< 0.6$ for the single-driver sources) under typical operating conditions. Similar values were found by Jain et al.~\cite{Jain_2022} at the SPIDER test bed. This raises the questions why less RF power per driver is coupled to a multi-driver plasma, when compared to a single-driver plasma and how this could be optimized. To answer this questions the self-consistent fluid-electromagnetic model is applied in both cases. The model is described in section~\ref{sec:model} and the calculated distributions of relevant quantities such as plasma density and electromagnetic fields are discussed in section~\ref{sec:results}. A conclusion as drawn from the model results is given in section~\ref{sec:conclusion}.

\section{Model}
\label{sec:model}

\begin{figure}[ht]
	\centering
	\includegraphics[width=0.9\textwidth]{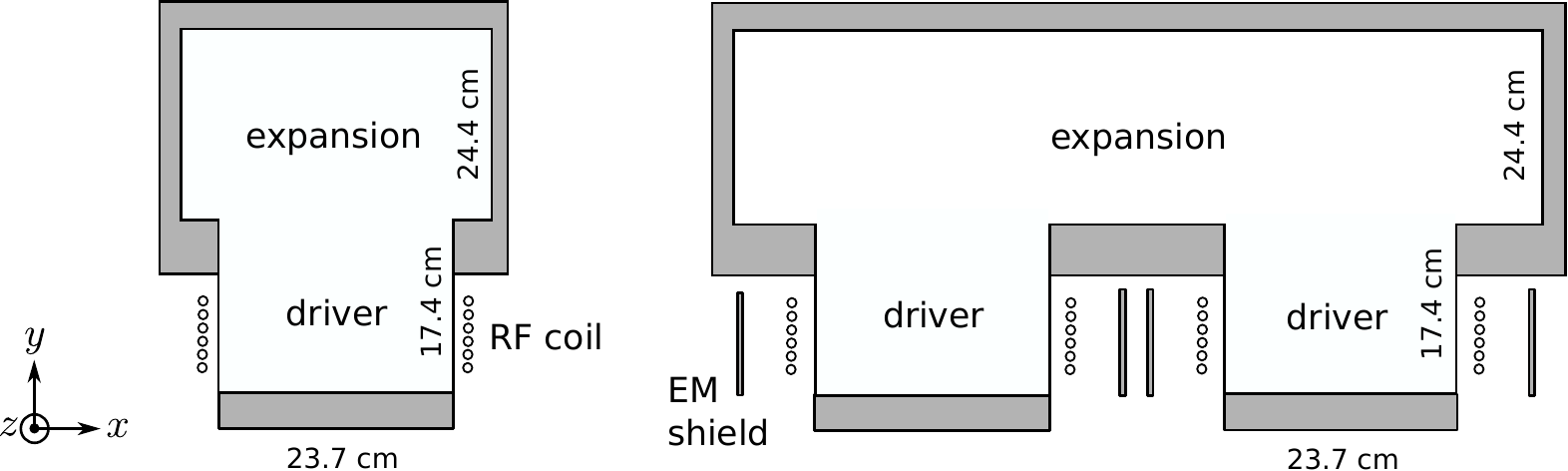}
	\caption{\label{fig:simulation_domain}Simulation domains corresponding to a horizontal cut for the single-driver (left) and multi-driver (right). The latter has electromagnetic shields surrounding each RF coil.}%
\end{figure}

To represent multi-driver setups, the cylindrical symmetry as assumed in our previous work~\cite{Zielke_2022} is not appropriate, wherefore the equations and the modeling domains used in this work are in 2D cartesian coordinates, as shown in figure~\ref{fig:simulation_domain}. 

It has been verified that the spatial plasma parameter profiles do not change significantly when using cartesian instead of cylindrical coordinates. The two different simulation domains represent horizontal cuts of the single- and multi-driver setups. Further improvements regarding the description of the RF power coupling are that the model in this work is fully time dependent, i.e.\ no time harmonic approximation is used, neither in the electron momentum balance, where also the viscous stress tensor and the Lorentz force are retained, nor in Maxwell's equations. This comes with the drawback, that the time steps during the solving process become very small (0.001 of the RF period, i.e.\ in the order of 1\,ns). For this reason the neutral molecules and atoms (which have a time-scale of around 0.1\,s) have to be described as a uniform background. The effect of neutral depletion is nevertheless accounted for by using the appropriate input values of $n_{\mathrm{H}_2} = 3\cdot 10^{19}\,\mathrm{m}^{-3}$ and $n_\mathrm{H} = 1.5\cdot 10^{19}\,\mathrm{m}^{-3}$, as calculated for a filling gas pressure of 0.3\,Pa from a version of the model, where the neutrals are treated as fluids~\cite{Zielke_2022}. All other equations and boundary conditions are as described in our previous work~\cite{Zielke_2022}, with the one further exception, that also the impact of the RF magnetic field on the electron heat flux is considered in this work, as described elsewhere~\cite{Zielke_2021_2}. The grey shaded surfaces in figure~\ref{fig:simulation_domain} represent perfect conductors, such as the driver and source backplates, as well as the electromagnetic shields. Here the electric component of the RF field is set to zero, i.e.~$E_z = 0$. The relation between the calculated RF coil current amplitude $I_\mathrm{coil}$, which is applied at each of the six coil windings (depicted as small circles left and right of each driver) and the tangential component of the magnetic RF field $B_\mathrm{RF,tan}$ is
\begin{equation}
	I_\mathrm{coil} = \frac{1}{\mu_0} \oint_\mathrm{L} B_\mathrm{RF,tan}\mathrm{d}L,
	\label{eq:I_coil}
\end{equation}
where the integral is along the circumference of each coil winding. The system is excited by an integral controller which ramps up $I_\mathrm{coil}$ to ensure that the absorbed power by the plasma is as specified by the user. This steady state is typically reached after approximately 70 RF cycles. The RF power transfer efficiency $\eta$, as the quantity which should be optimized, calculates the ratio of power absorbed by the plasma $P_\mathrm{plasma}$ and total delivered generator power $P_\mathrm{RF}$,
\begin{equation}
	\eta = \frac{P_\mathrm{plasma}}{P_\mathrm{RF}} = \frac{P_\mathrm{plasma}}{P_\mathrm{plasma} + \frac{1}{2}R_\mathrm{net}I_\mathrm{coil}^2}.
	\label{eq:eta}
\end{equation}
Herein it is convenient to express the losses in the RF network as a product of a network resistance $R_\mathrm{net}$ and the square of the RF coil current $I_\mathrm{coil}$. The network resistance is not in the focus of this study, wherefore it is a model input. Its value of $R_\mathrm{net,SD} = 0.6\,\Omega$ has been determined experimentally for the single-driver ion source at the BUG test bed~\cite{Zielke_2021}. To keep things as simple as possible the driver diameters in the multi-driver model are kept the same as in the single-driver (see figure~\ref{fig:simulation_domain}) and the network resistance is simply doubled, i.e.\ $R_\mathrm{net,MD} = 1.2\,\Omega$ is used. Note that the experimentally obtained value for the upper two drivers at ELISE is $1.28\,\Omega$ and thus slightly larger. The deviation is most likely caused by eddy current losses in the electromagnetic shields and the fact that the drivers in ELISE are slightly larger in the experimental setup (driver inner diameter of 27.6\,cm instead of 23.7\,cm). This entails a larger Faraday shield surface, where eddy currents are driven and a longer RF coil.

\section{Results}
\label{sec:results}	

\subsection{Single- and multi-driver setup without electromagnetic shields}

\begin{figure}[ht]
	\centering
	\includegraphics[width=0.8\textwidth]{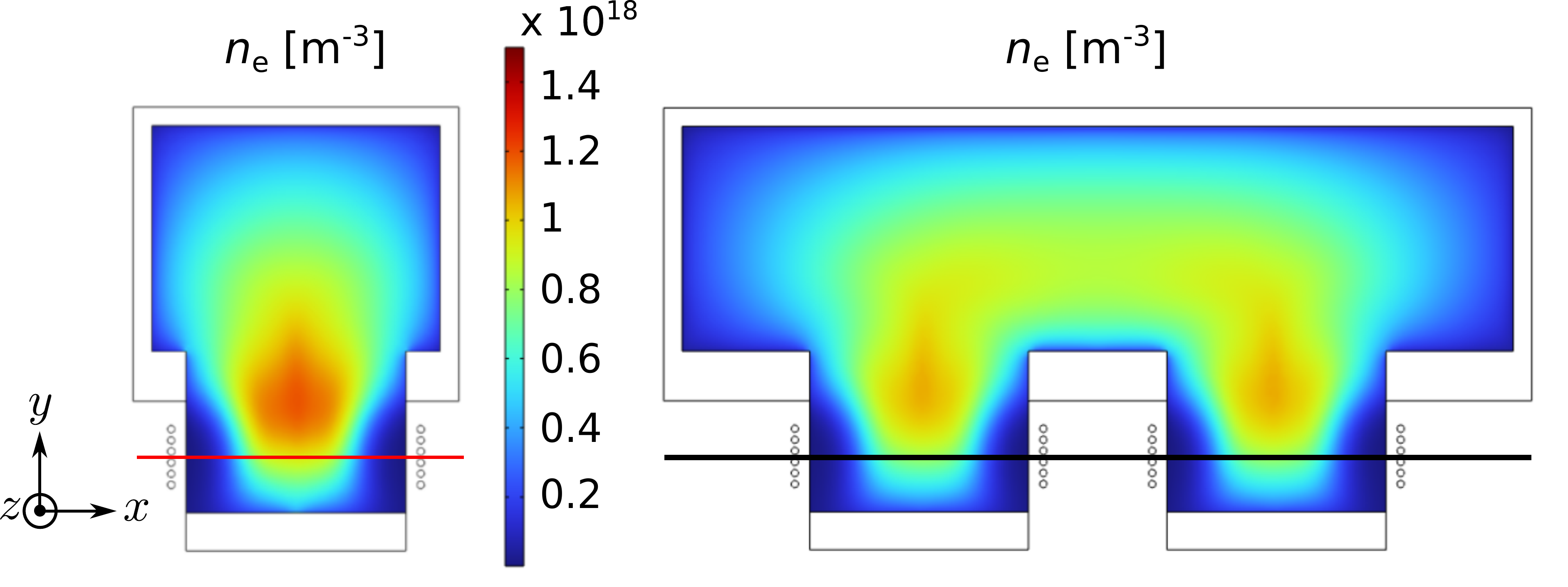}
	\caption{\label{fig:comparison_ne_single_multiple_driver}Comparison of calculated electron density profiles in a single- and multi-driver setup (without electromagnetic shields around the RF coils) for an absorbed power of 25\,kW by the plasma in each driver.}
\end{figure}

Figure~\ref{fig:comparison_ne_single_multiple_driver} shows a comparison of the spatial profile of the electron density in the single- and multi-driver setup, averaged over one RF period in steady state, when 25\,kW of power are absorbed in each driver plasma. In this first comparison, no electromagnetic shields are used in the multi-driver setup.
The shape shows the typical depletion of electrons near the RF coil (see dark blue regions), where the RF Lorentz force tends to push the plasma away from the RF coils. Power absorption profiles (not shown) and plasma parameter profiles are virtually the same in each driver of both setups. Using in the multi-driver case the doubled plasma power and network resistance in equation~(\ref{eq:eta}) it can be shown that for $\eta$ to remain unchanged, it is necessary that also $I_\mathrm{coil}$ does not change. However, the model shows that this is not the case, i.e.\ the respective RF coil currents to produce these plasmas are $I_{\mathrm{coil,SD}} = 263\,$A for the single-driver and $I_{\mathrm{coil,MD}} = 284\,$A for the multi-driver setup.
\begin{figure}[h]
	\centering
	\includegraphics[width=0.7\linewidth]{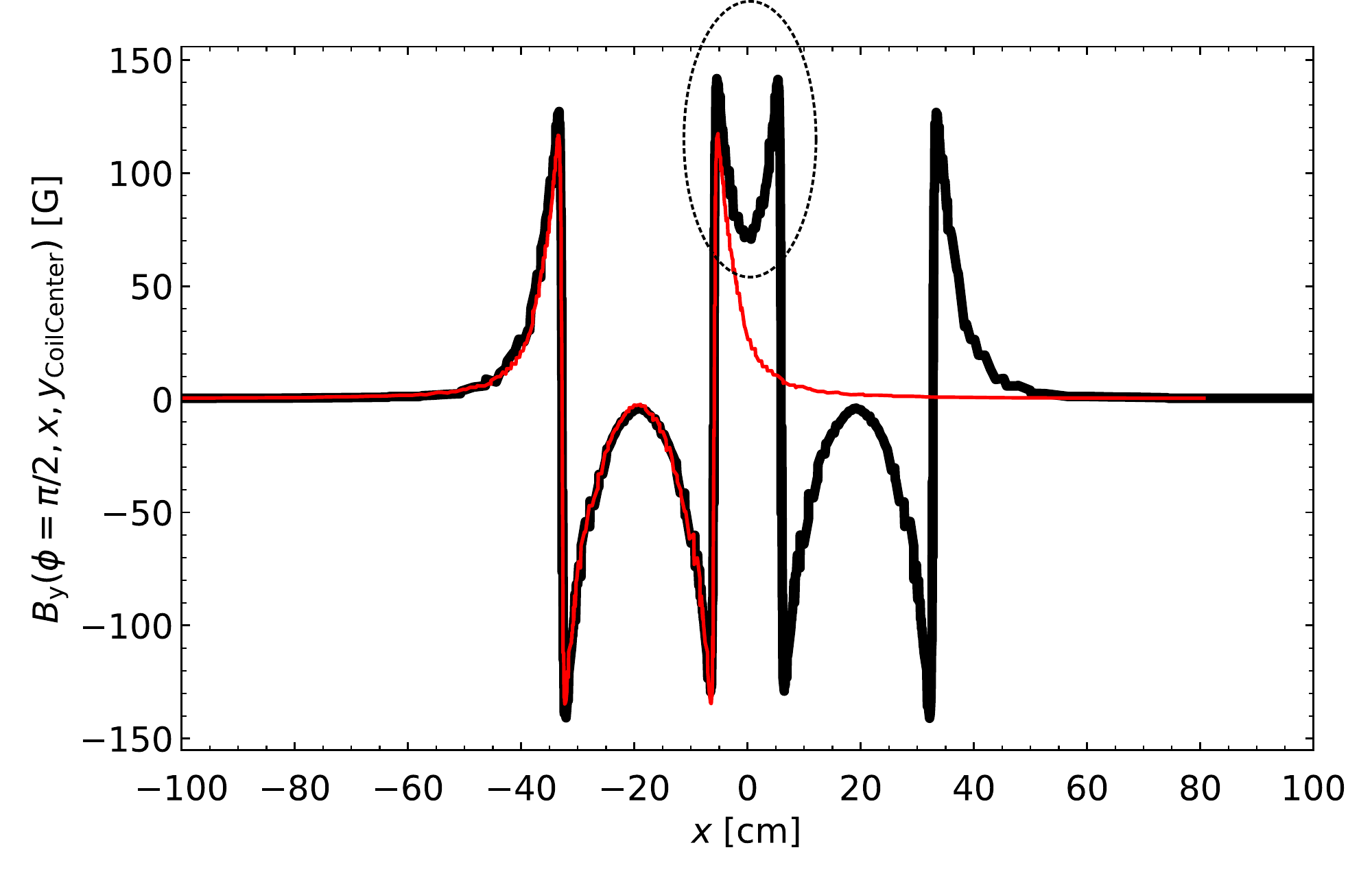}
	\caption{\label{fig:comparison_By}Comparison of the $y$-component of the RF magnetic field $B_y$ at a phase of $\pi/2$ along the lines as depicted in figure~\ref{fig:comparison_ne_single_multiple_driver}.}%
\end{figure}
Hence the coil current is increased by around 8\%. As a result, $\eta_\mathrm{SD} = 55\% > \eta_\mathrm{MD} = 50\%$. The reason for the lower $\eta_\mathrm{MD}$ is electromagnetic mutual coupling between the two drivers. This can be seen exemplarily in figure~\ref{fig:comparison_By}, where the $y$-components of the magnetic RF fields (at a phase $\phi = \pi/2$) along the lines as depicted in figure~\ref{fig:comparison_ne_single_multiple_driver} are compared. For better comparability, the field from the single-driver setup has been shifted onto the left driver of the multi-driver setup. As can be seen, $B_y$ (representative for $B_\mathrm{RF,tan}$) at the positions of the coils is larger in the case of the multi-driver setup. According to equation~(\ref{eq:I_coil}) this results in a larger $I_\mathrm{coil}$ and consequently a lower $\eta$.

\subsection{Effect of electromagnetic shields}

\begin{figure}[ht]
	\centering
	\includegraphics[width=0.9\linewidth]{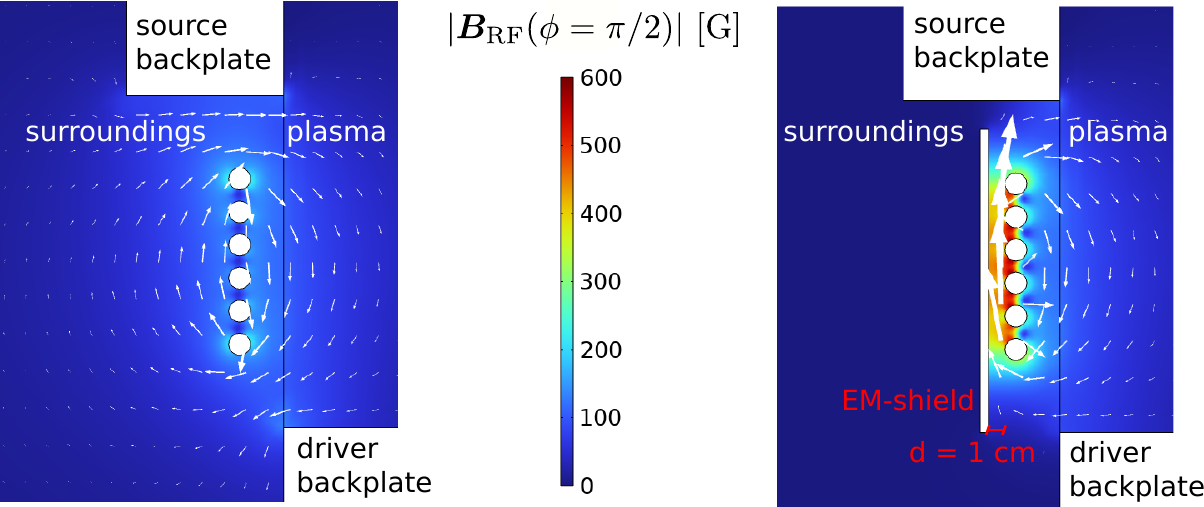}
	\caption{\label{fig:2DBFieldDistribution}Distributions of the magnetic RF field for the case without EMS (left) and with EMS (right). A close distance of 1\,cm is used to emphasize the shielding effect.}%
\end{figure}

To avoid mutual coupling between individual drivers in multi-driver setups conductive cylindrical electromagnetic shields (EMS) are wrapped around each coil and attached to the grounded source backplates~\cite{Kraus_2012}, as shown in figure~\ref{fig:CAD_drivers}. To investigate how the RF power coupling is affected by the EMS, the simulation domain including the EMS is used, as shown in figure~\ref{fig:simulation_domain}, right. As a boundary condition at the conducting electromagnetic shield, $E_z = 0$ is used. Figure~\ref{fig:2DBFieldDistribution} shows the resulting distribution of the RF magnetic field amplitude, illustrated for an exemplary distance of 1\,cm between RF coil and EMS.

The white arrows represent the direction and relative strength of the RF magnetic field amplitude. On the left, no EMS is present, wherefore the magnetic RF field amplitude around each RF coil winding is uniform and comparatively low. In contrast to that, on the right, the distribution is highly anisotropic. Consequently the current that has to be applied at the RF coil windings to reach the same absorbed plasma power is larger in this case. 

\begin{figure}[ht]
	\centering
	\includegraphics[width=0.5\linewidth]{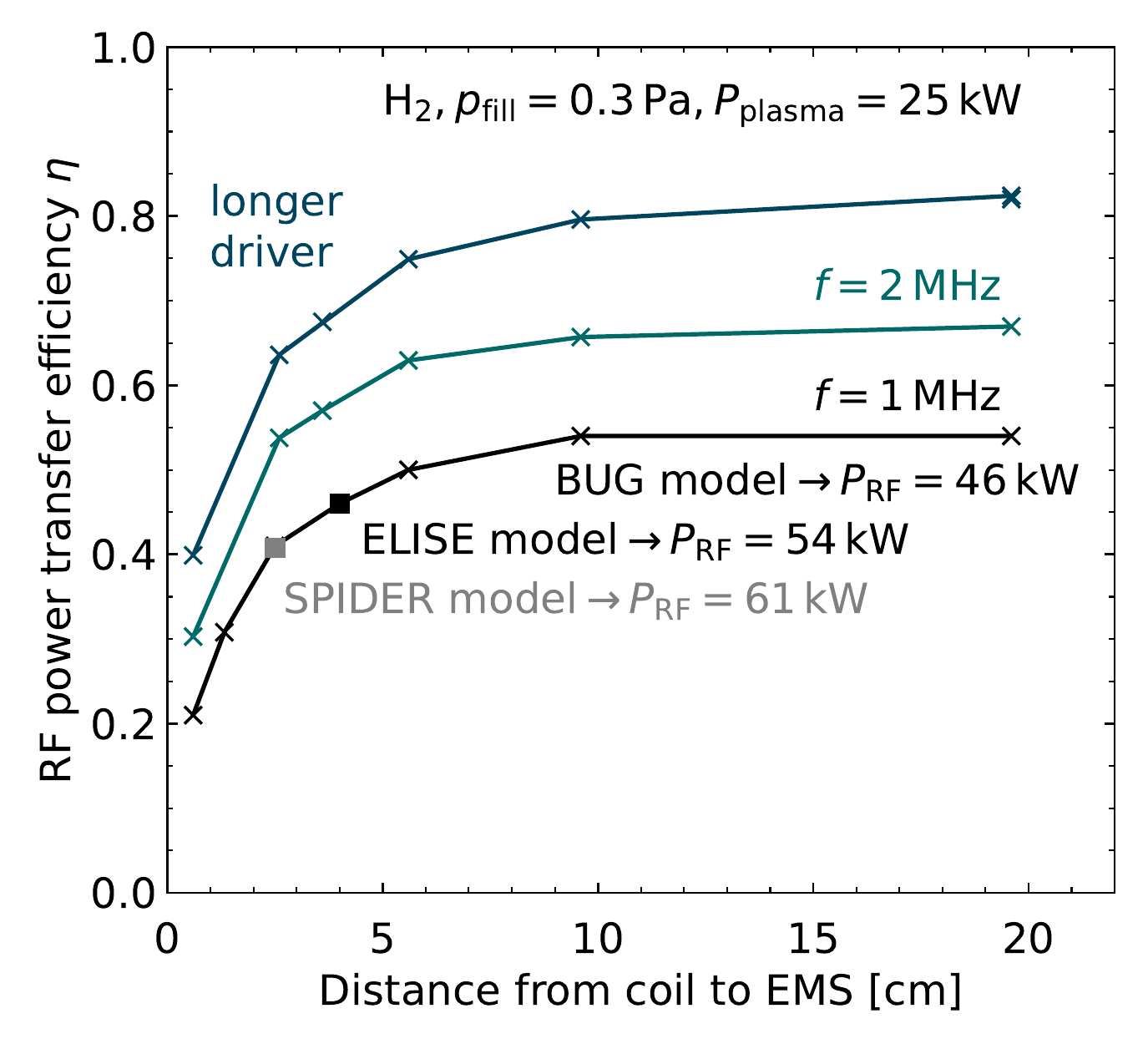}
	\caption{\label{fig:eta_distance_coil_emshield}RF power transfer efficiency $\eta$ as a function of the distance between RF coil and EMS in a single-driver setup.}%
\end{figure}

For a distance of 4\,cm, as in the ELISE ion source, the model calculates an RF power transfer efficiency of 0.46. This is larger than the experimentally obtained value, which is $0.3 \pm 0.1$ without applying a magnetic filter field. Note that the latter is typically applied in the expansion to decrease the electron temperature and density, which helps to reduce the number of co-extracted electrons as well as the destruction processes of negative hydrogen ions~\cite{Fantz_2014}. The overestimation of $\eta$ by the model could be caused by 3D effects, as e.g.~conductive support structure to hold the EMS, which decreases the effective distance, as shown in figure~\ref{fig:CAD_drivers}. The distance $d$ between the RF coil and the EMS is a crucial parameter. Thus the single-driver setup is used to quantify the impact of $d$ on the RF power transfer efficiency isolated from any other effects encountered in multi-driver sources. The resulting behavior is shown in figure~\ref{fig:eta_distance_coil_emshield}.

As the solid black line shows, for large distances $\eta$ approaches the value of a setup, where no EMS is present, i.e.\ around 0.55. However, when the distance is decreased $\eta$ decreases non-linearly to around 0.2 at 1\,cm. In this extreme case, a generator output power of 125\,kW per driver would be necessary to sustain a plasma with an absorbed power of 25\,kW. The black square represents the multi-driver model of the ELISE ion source with $d = 4\,$cm. Note that it almost perfectly fits onto the curve. This justifies using a model with a single driver to investigate the distance effect and also shows that this is the main effect, which affects the RF power coupling. As another example the SPIDER ion source, where the distance equals 2.5\,cm is also plotted onto the curve as a grey square. In the two cases RF generator powers of 54\,kW and 61\,kW per driver are required, respectively.

In order to assess the optimization possibilities for the RF power coupling, two measures as proposed in an earlier work~\cite{Zielke_2022_2} are applied in the multi-driver model with EMS. The two resulting curves show that both measures, i.e.~applying 2\,MHz and increasing the axial driver length from 17\,cm to 30\,cm are beneficial. For the case of ELISE, where $d = 4\,$cm, $\eta$ could be increased to around 0.57 and 0.68, respectively. Furthermore, when both measures are combined at the same time, the model predicts that an RF power transfer efficiency of 0.74 is reachable. Nevertheless, the negative effect of the EMS on the RF power transfer efficiency is persistent, as the model predicts values of up to 0.9 without them~\cite{Zielke_2022_2}.

\section{Conclusion}
\label{sec:conclusion}

A state-of-the-art 2D fluid-electromagnetic model has been further developed and applied to self-consistently describe the power coupling in fusion-relevant single- and multi-driver RF ion sources. At the same absorbed plasma power per driver, the numerically calculated plasma parameters in single- and multi-driver setups coincide. However, a slightly decreased RF power transfer efficiency is found in multi-driver setups, even without electromagnetic shields present. The cause for this decrease is mutual electromagnetic coupling, i.e.\ the presence of a second driver in close vicinity of the first one changes the distribution of the magnetic RF field. Moreover, the electromagnetic shields present in multi-driver RF ion sources are identified to considerably lower the RF power transfer efficiency. This is caused by a drastic change of the magnetic RF field distribution, which is more nonuniform around the individual coil windings when an electromagnetic shield is present. The nonuniform distribution leads to a larger RF coil current and consequently a lower RF power transfer efficiency. The decrease in RF power transfer efficiency is found to be highly nonlinear with decreasing distance between the coil and electromagnetic shield. Measures to increase the power transfer to the plasma as already found in previous works could be applied in multi-driver ion sources as well. By enlarging the axial driver length from 17\,cm to 30\,cm and increasing the driving frequency from 1\,MHz to 2\,MHz at the ELISE ion source the model predicts an increase of the RF power transfer efficiency from 0.46 to 0.74.

\acknowledgments

This work has been carried out within the framework of the EUROfusion Consortium and has received funding from the Euratom research and training programme 2014-2018 and 2019-2020 under grant agreement No 633053. The views and opinions expressed herein do not necessarily reflect those of the European Commission.

\noindent{This work has been carried out within the framework of the EUROfusion Consortium, funded by the European Union via the Euratom Research and Training Programme (Grant Agreement No 101052200 — EUROfusion). Views and opinions expressed are however those of the author(s) only and do not necessarily reflect those of the European Union or the European Commission. Neither the European Union nor the European Commission can be held responsible for them.}



\begin{thebibliography}{99}

\bibitem{ITER}
\emph{www.iter.org}

\bibitem{Kraus_2012}
W.~Kraus et al.~\emph{Re.~Sci.~Instrum.} {\bf 83, 2} (2012) 02B104

\bibitem{Heinemann_2017}
B.~Heinemann et al.~\emph{New J.~Phys.} {\bf 19, 1} (2017) 015001

\bibitem{Fantz_2014}
U.~Fantz et al.~\emph{Plasma Sources Sci.~Technol.} {\bf 23, 4} (2014) 044002

\bibitem{Fantz_2017}
U. Fantz et al.~\emph{Nucl.~Fusion} {\bf 57, 11} (2017) 116007

\bibitem{Hemsworth_2017}
R.~Hemsworth et al.~\emph{New J.~Phys.} {\bf 19, 2} (2017) 025005

\bibitem{Zielke_2021}
D.~Zielke et al.~\emph{J.~Phys.~D} {\bf 54, 15} (155202) 155202

\bibitem{Zielke_2021_2}
D.~Zielke et al.~\emph{Plasma Sources Sci.~Technol.} {\bf 30, 6} (2021) 065011

\bibitem{Zielke_2022}
D.~Zielke et al.~\emph{Plasma Sources Sci.~Technol.} {\bf 31, 3} (2022) 035019

\bibitem{Jain_2022}
P. Jain et al.~\emph{Plasma Phys.~Control.~Fusion} {\bf 64, 9} (2022) 095018

\bibitem{Zielke_2022_2}
D.~Zielke et al.~\emph{submitted to Nucl.~Fusion} (2022)

\bibitem{Kraus_2001}
W. Kraus et al.~\emph{Fusion Eng.~Des.} {\bf 56-57} (2001) 499-503




% Please avoid comments such as "For a review'', "For some examples",
% "and references therein" or move them in the text. In general,
% please leave only references in the bibliography and move all
% accessory text in footnotes.

% Also, please have only one work for each \bibitem.


\end{thebibliography}
\end{document}